\documentstyle[amssymb,preprint,aps]{revtex}

\begin{document}
\draft
\title{Quantum dissipative chaos in the statistics of excitation numbers}
\author{Gagik Yu. Kryuchkyan and Suren B. Manvelyan}
\address{Yerevan State University, Manookyan 1, Yerevan, 375049, Armenia\\
and\\
Institute for Physical Research, National Academy of Sciences,\\
Ashtarak-2, 378410, Armenia}
\maketitle

\begin{abstract}
A quantum manifestation of chaotic classical dynamics is found in the
framework of oscillatory numbers statistics for the model of nonlinear
dissipative oscillator. It is shown by numerical simulation of an ensemble
of quantum trajectories that the probability distributions and variances of
oscillatory number states are strongly transformed in the order-to-chaos
transition. A nonclassical, sub-Poissonian statistics of oscillatory
excitation numbers is established for chaotic dissipative dynamics in the
framework of Fano factor and Wigner functions. It is proposed to use these
results in experimental studies and tests of the quantum dissipative chaos.
\end{abstract}


The quantum effects inherent in the chaotic behavior of classical systems is
an interesting field pertaining to many problems of fundamental interest 
\cite{1}. Though these problems have been posed long ago, they are still of
great interest. The study of quantum dynamics of isolated or so-called
Hamiltonian systems, the classical counterparts of which are chaotic, has a
long history. In the majority of studies the attention has been focused on
static properties such as spectral statistics of energy levels and the
transition probabilities between eigenstates of the system. A variety of
studies have been also carried out for understanding the features of
time-dependent chaotic systems \cite{2}. By contrast to that, few works
dealt with to investigations of the quantum chaos for open nonlinear
systems. The early studies of open chaotic systems date back from the work
by Ott et al. \cite{3}, and papers of Graham and Dittrich \cite{4}, where
the authors have analyzed the kicked rotor and similar discrete-time systems
. The investigations of quantum chaotic systems are distinctly connected
with the quantum-classical correspondence problem in general and with
environment induced decoherence and dissipation in particular. Recently this
topic has been in the focus of theoretical investigations. As part of these
studies it has been recognized by Zurek and Paz \cite{5}\ and in later works 
\cite{6} that the decoherence has rather unique properties for systems the
classical analogues of which are chaotic ones. Among the criteria suggested
for definition of chaos in open quantum systems one is to separate those
based on entropy production and Wigner functions \cite{4}- \cite{6}. From
experimental viewpoint, the observation of dissipative effects and
environment induced decoherence of dynamically localized states in the
quantum delta-kicked rotor is made using the gas of ultracold cesium atoms
in a magneto-optical trap subjected to a pulsed standing wave \cite{7,8}.
Recently, new problems of chaotic motion were studied in an experimental
scheme with ultra-cold atoms in magneto-optical double-well potential \cite
{9}. In spite of these important developments in the investigation of chaos
for open quantum systems, there are still many open questions, and there is
a clear need in new models that admit experimental verification, as well as
comparatively more simple physical criterions for testing the dissipative
quantum chaos.

The first purpose of this Letter is to investigate the order-to-chaos
transition at the level of statistics of elementary excitations for the
quantum model of nonlinear oscillator. We show below that the distributions
of oscillatory occupation numbers can be used to distinguish between the
ordered and chaotic quantum dissipative dynamics. The need in realization of
this study is to have a proper quantum model showing both the regular and
chaotic dynamics in the classical limit. This aim in view we propose a
nonlinear oscillator driven by two forces at different frequencies. This
model was proposed to study the quantum stochastic resonance and quantum
chaos in our previous papers \cite{10}, where it was shown in details that
the model is suited to verification in experiments. Our second purpose is to
identify the\ kind of statistics of oscillatory number states taking place
in the quantum chaos. Our central result here is that the nonclassical,
sub-Poissonian statistics can be realized for chaotic dynamics of the system
under consideration.

The open quantum systems are usually studied in the framework of reduced
density matrix obtained by tracing it over the degrees of freedom of a
reservoir. The evolution of the system of interest is governed by the
following master equation for the reduced density matrix in the interaction
picture

\begin{equation}
\frac{\partial \rho }{\partial t}=-\frac{i}{\hbar }\left[ H,\rho \right]
+\sum_{i=1,2}\left( L_{i}\rho L_{i}^{+}-\frac{1}{2}L_{i}^{+}L_{i}\rho -\frac{%
1}{2}\rho L_{i}^{+}L_{i}\right) ,  \label{mastereq}
\end{equation}
where the Hamiltonian \ \ \ \ \ \ \ 
\begin{equation}
H=\ \hbar \Delta a^{+}a+\hbar \left[ \left( \Omega _{1}+\Omega _{2}\exp
\left( -i\delta t\right) \right) a^{+}+\left( \Omega _{1}^{\ast }+\Omega
_{2}^{\ast }\exp \left( i\delta t\right) \right) a\right] +\hbar \chi
(a^{+}a)^{2}  \label{hamiltonian}
\end{equation}
describes an anharmonic oscillator with oscillatory frequency $\omega _{0}$
driven by two periodic forces at frequencies $\omega _{1}$ and $\omega _{2}$%
. The couplings with two driving forces are given by Rabi frequencies $%
\Omega _{1}$ and $\Omega _{2}$, and $\chi $ is the strength of
anharmonicity. Here$\ \Delta =\omega _{0}-$ $\omega _{1}$\ is the detuning, $%
\delta =\omega _{2}-$ $\omega _{1}$ is the difference between driving
frequencies, that plays the role of modulation frequency in the interaction
picture, and $a,$ $a^{+}$ are boson annihilation and creation operators. The
last terms in Eq. (\ref{mastereq}) pertain to the influence of the
environment induced diffusion. $L_{i}$ are the Lindblad operators: $L_{1}=%
\sqrt{\left( N+1\right) \gamma }a,\;L_{2}=\sqrt{N\gamma }a^{+}$, where $%
\gamma $ is the effective decay rate of the dissipation process and $N$
denotes the mean number of quanta of the heat bath. We have followed the
standard approach \cite{11} to dissipative quantum dynamics in the range of
weak coupling of system with the reservoir under the condition: $\gamma \ll
k_{B}T/\hbar ,$ where $k_{B}T$ is the Boltzman's constant times temperature.
Equation (\ref{mastereq}) is obtained in both the rotating wave and Markov
approximations, without regard for the driving-induced noise effects \cite
{12}.

For $\Omega _{2}=0$ this equation describes the single driven, dissipative
anharmonic oscillator, which is a well-known and archetypal model in
nonlinear physics \cite{13}. In case of double driven oscillator ($\Omega
_{2}\neq 0$), the Hamiltonian (\ref{hamiltonian}) is explicitly
time-dependent and the system exhibits regions of regular and chaotic
motion. In the classical limit, the corresponding equation of motion for the
dimensionless amplitude $\alpha (t)=\left\langle a(t)\right\rangle $ has the
form 
\begin{equation}
\frac{d}{dt}\alpha =-\frac{1}{2}\gamma \alpha -i\left( \Delta +\chi
(1+2\left| \alpha \right| ^{2})\right) \alpha -i\left( \Omega _{1}+\Omega
_{2}\exp \left( -i\delta t\right) \right) .  \label{clas}
\end{equation}
\qquad 

It should be noted that our model corresponds to a modified model of Duffing
oscillator. Indeed, it is easy to demonstrate using the result of \cite{14}
that Eq.(3) is the rotating-wave approximation of the equation of Duffing
oscillator driven by two periodic forces. At first, we study the
order-to-chaos transition of the system in question using a constant phase
map in the phase space.\ Our numerical analysis of Eq.(3) in $(X,Y)$ plane ($%
X=%
\mathop{\rm Re}%
\alpha $, $Y=%
\mathop{\rm Im}%
\alpha )$ shows that the classical dynamics of the system is regular in
domains of small and large values of modulation frequency, i.e. $\delta \ll
\gamma $ and $\delta \gg \gamma $, and also when one of the perturbation
forces is much greater than the other: $\Omega _{1}\ll \Omega _{2}$ or $%
\Omega _{2}\ll \Omega _{1}.$ The dynamics is chaotic in the range of
parameters $\delta \gtrsim \gamma $ and\ $\Omega _{1}\simeq \Omega _{2}$,
where the classical strange attractors for the Poincar\'{e} section are
realized (see Fig.1).

\ The proposed model does not allow an analytical description. Our numerical
analysis is based on quantum state diffusion approach that represents the
reduced density operator by the mean over the projectors onto the stochastic
states $\left| \Psi _{\xi }\right\rangle $ of the ensemble: $\rho
(t)=M\left( \left| \Psi _{\xi }\right\rangle \left\langle \Psi _{\xi
}\right| \right) $, where $M$ denotes the ensemble averaging. The
corresponding equation of motion is

\begin{eqnarray}
\left| d\Psi _{\xi }\right\rangle  &=&-\frac{i}{\hbar }H\left| \Psi _{\xi
}\right\rangle dt-  \label{QSD} \\
&&\frac{1}{2}\sum_{i=1,2}\left( L_{i}^{+}L_{i}-2\left\langle
L_{i}^{+}\right\rangle L_{i}+\left\langle L_{i}\right\rangle \left\langle
L_{i}^{+}\right\rangle \right) \left| \Psi _{\xi }\right\rangle
dt+\sum_{i=1,2}\left( L_{i}-\left\langle L_{i}\right\rangle \right) \left|
\Psi _{\xi }\right\rangle d\xi _{i},  \nonumber
\end{eqnarray}
where $\xi $ indicates the dependence on the stochastic process, the complex
Wiener variables $d\xi _{i}$ satisfy the fundamental properties $M\left(
d\xi _{i}\right) =0,\;M\left( d\xi _{i}d\xi _{j}\right) =0,\;M\left( d\xi
_{i}d\xi _{j}^{\ast }\right) =\delta _{ij}dt,$ and the expectation value $%
\left\langle L_{i}\right\rangle =\left\langle \Psi _{\xi }\left|
L_{i}\right| \Psi _{\xi }\right\rangle $.

Numerical studies of the oscillatory mean excitations number $\left\langle
n\right\rangle =M(\left\langle \Psi _{\xi }\right| a^{+}a\left| \Psi _{\xi
}\right\rangle )$ show that in both cases of regular and chaotic dynamics
this quantity exhibits a periodic time-dependent behavior that is
approximately sinusoidal with a period of $2\pi /\delta .$ We see that the
quantum manifestation of chaotic dissipative dynamics is not evident on the
mean oscillatory number. We will study the macroscopic quantum effects
assisting to chaotic behavior by consideration of both the probability
distribution of oscillatory excitation numbers $P_{n}=\left\langle n\right|
\rho \left| n\right\rangle ,$ where $\left| n\right\rangle $ is the number
states, and the Fano factor which describes the excitation number
uncertainty, normalized to the level of fluctuations for coherent states, i.
e. $F=\left\langle \left( \Delta n\right) ^{2}\right\rangle /\left\langle
n\right\rangle ,$ $\left\langle \left( \Delta n\right) ^{2}\right\rangle
=\left\langle \left( a^{+}a\right) ^{2}\right\rangle -\left\langle
a^{+}a\right\rangle ^{2}$. This investigation will be complemented by
testing the quantum chaos in phase space with the help of numerical
calculations of the Wigner function.

To study the pure quantum effects we focus on the cases of very low
reservoir's temperatures which, however, ought to be still larger than the
characteristic temperature $T\gg T_{ch}=\hbar \gamma /k_{B}.$ This
restriction implies that dissipative effects can be described
selfconsistently in the frame of the Lindblad equation (\ref{mastereq}). For
clarity, in our numerical calculation we choose the mean number of reservoir
photons $N=\left( e^{\hbar \omega /k_{B}T}-1\right) ^{-1}$ equaled to $%
N=0.002.$ Note here that for $N\ll 1$ the above restriction is valid for the
majority of problems of quantum optics and, particularly, for the scheme
involving a trapped electron. In this scheme, $N=0.002$ for the microwave
spectral range corresponds to $T\simeq 0.16$ K, while $T_{ch}=10^{-9}$ K, as 
$\gamma \sim 10^{2}$ $s^{-1}$. The detailed discussion of possible
experimental realizations is postponed to the end of the Letter. .

Let us first consider the quantities of interest for the region of
classically regular behavior with parameters: $\chi /\gamma =0.1$, $\Delta
/\gamma =-15$, $\Omega _{1}/\gamma =27$, $\Omega _{2}/\gamma =35$, and $%
\delta /\gamma =5$. In Fig.2 the Wigner function is shown at the fixed
moments $t_{n}=[7.13+(2\pi /\delta )n]\gamma ^{-1},$ $(n=0,1,2,...)$
exceeding the transient time. We find that the Wigner function located
around the point $X=0,$ $Y=-10$ and its contour-plot has a narrow crescent
form with the origin of phase space as its centrum. The radial squeezing
that reproduces the known property of the anharmonic oscillator model to
generate the excitation number squeezing is also clearly seen in the figure.
The important novelty here is that the radial squeezing effect in this model
is much stronger, than an analogous one for the model of single driven
anharmonic oscillator \cite{13}. Below we will quantitatively demonstrate
this by analyzing the Fano factor. Another peculiarity here is that the
Wigner function is nonstationary. As the calculations show, during the
modulation period $2\pi /\delta $ it is rotated around the origin of the
phase space. In Fig.3 (curve 1) one can see the time evolution of the Fano
factor, which shows the formation of nonclassical sub-Poissonian statistics (%
$\left\langle \left( \Delta n\right) ^{2}\right\rangle <\left\langle
n\right\rangle $) for time intervals exceeding the transient time. The Fano
factor reaches its minimum $F_{\min }\simeq 0.12$ and maximum $\ F_{\max
}\simeq 0.65$ values in definite time intervals. One can account for
surprisingly high sub-Poissonian statistics only by the quantum nature of
oscillatory excitations under the influence of two driving forces. Indeed,
in case of $\Omega _{2}=0$ \ we have $F\simeq 0.35$ for the same parameters
as those in Fig.3 (curve 1).

We are now in a position to study the rise of quantum chaos, which is
expected to manifest itself as crucial changes in above results at the
passage into the classically chaotic operational regime, with parameter
values: $\chi /\gamma =0.1,$ $\Delta /\gamma =-15,$ $\Omega _{1}/\gamma
=\Omega _{2}/\gamma =27,$ and $\delta /\gamma =5.$ In this range the
oscillatory mean excitations number oscillates between $\left\langle
n\right\rangle =70\div 130.$ Now consider the behavior of the Fano factor.
Its time evolution is shown in Fig.3 (curve 2). Surprisingly, the
excitation-number fluctuations are also squeezed below the coherent level
under the considered chaotic regime. However, opposite to the previous
regular regime, the excitation number exhibits both the sub-Poissonian ($F<1$%
) and super-Poissonian ($F>1$)\ statistics, that are alternating in definite
time intervals. The minimum and maximum values of $F$ in time intervals
during one modulation period are equal to $F_{\min }\simeq 0.30$ and $%
F_{\max }\simeq 1.98$. Thus, Fig.3 shows the drastic difference between the
behavior of Fano factor for regular and chaotic dynamics.

It is tempting to explain the emergence of nonclassical sub-Poissonian
statistics in the double driven nonlinear oscillator at the transition from
regular to chaotic dynamics using the phase space symmetry properties of the
Wigner function. The results of ensemble-averaged numerical calculations of
the contour-plot of Wigner function at fixed time intervals $%
t_{n}=[6.96+(2\pi /\delta )n]\gamma ^{-1},$ $(n=0,1,2,...)$ is shown in
Fig.4. It is seen that the contour-plot for chaotic motion still has the
radial squeezed form (see Fig.4). This result takes place for $%
t_{n}=[6.96+(2\pi /\delta )n]\gamma ^{-1},$ $(n=0,1,2,...),$ at which the
Fano factor reaches its minimum value $F_{\min }\simeq 0.30$. In the next
time intervals during the period of modulation, the level of excitation
number fluctuations increases, and as a result the radial squeezing in
contour-plot decreases. It is easy to see that the contour-plot is generally
similar to the correspondent classical Poincar\'{e} section (Fig.1).

In the search for a criterion of quantum chaos that is more promising and
easily attainable in experiments, we consider the probability distribution
of oscillatory excitation numbers $P_{n}=\left\langle n\left| \rho \right|
n\right\rangle $. We give in Fig.5 the results for both regular (a) and
chaotic (b) regimes at two time moments corresponding to $F_{\min }$ (curve
1) and $F_{\max }$ (curve 2). One can conclude from the comparison of these
figures that the probability distributions $P_{n}$ are strongly transformed
at the order-to-chaos transition. While $P_{n}$ for regular dynamics is
clearly bell-shaped and localized in narrow intervals of oscillatory
numbers, the distribution for chaotic dynamics is flat-topped with
oscillatory numbers from $n=0$ to $n_{\max }\simeq 200.$ Moreover, the shape
of distributions changes irregularly in time during the period $2\pi /\delta 
$. Especially typical for chaotic motion is the result shown in Fig. 5(b)
(curve 2), where the probability distribution is almost equally probable.

We also studied the reliability of our results depending on noise intensity.
We found that the results had not changed visibly in the range of $0\leq
N\lesssim 0.05$. For example, $F_{N=0.05}-F_{N=0}\sim 0.1$. What concerns
the Wigner function, it is only slightly smeared in the phase space under
the influence of such an amount of noise.

We emphasize, that the number of possible experimental schemes demonstrating
the proposed model is rather large. One of those is that of a single
relativistic electron in Penning trap, which is a realization of anharmonic
oscillator as was predicted theoretically by Kaplan \cite{15} and
experimentally realized by Gabrielse and co-authors\cite{16}. In the
presence of two microwave electromagnetic fields this system gives an
example of double driven anharmonic oscillator and may be used for
demonstration of quantum dissipative chaos. This system is governed by Eq. (%
\ref{mastereq}), where the operators $a$ and $a^{+}$ describe the cyclotron
quantized motion, Rabi frequencies $\Omega _{1}$ and $\Omega _{2}$
characterize the amplitudes of the microwave driving fields, $\chi $ is the
strength of the anharmonicity due to relativistic effects, and $\gamma $ is
the spontaneous decay rate of the cyclotron motion.

In conclusion, we should like to summarize that the quantum-statistical
effects that accompany the chaotic dynamics have been identified. These
results were obtained for the model of dissipative anharmonic oscillator,
which has been proposed for studies of quantum chaos. It was demonstrated
that the oscillatory excitation numbers statistics could be used for the
diagnostic of quantum chaos. Indeed, we have shown that such measurable
quantities as the Fano factor and probability distributions of number states
are drastically changed at the order-to-chaos transition. But perhaps even
more intriguing are the results that nonclassical, sub-Poissonian statistics
of oscillatory number states is realized for the chaotic dissipative
dynamics. The results of our numerical work were obtained under conditions
of strong anharmonicity $\chi /\gamma \lesssim 1,$ for the value $\chi
/\gamma =0.1$, which is close to those actually achieved in the experiments
with trapped relativistic electron. We believe that the results obtained are
applicable to more general quantum systems, the classical analogues of which
exhibit chaotic dynamics.

\begin{center}
{\bf FIGURE CAPTIONS}
\end{center}

Fig. 1. The Poincar\'{e} section (approximately 20000 points) of a solution
to (3) plotted at times of the constant phase $t_{n}=6.96+(2\pi /\delta )n$, 
$n=0,1,2,..$. The dimensionless parameters are in the range of chaos: $\chi
/\gamma =0.1,$ $\Delta /\gamma =-15,$ $\ \Omega _{1}/\gamma =\Omega
_{2}/\gamma =27,$ $\delta /\gamma =5.$

Fig. 2 The Wigner function for the regular regime averaged over 3000
trajectories.

Fig. 3. The Fano factor versus dimensionless time for the regular (curve 1)
and chaotic (curve 2) regimes. The parameters are: $\chi /\gamma =0.1,$ $%
\Delta /\gamma =-15,$ $\ \Omega _{1}/\gamma =27,$ $\delta /\gamma =5,$ and $%
\ \Omega _{2}/\gamma =35$ (curve 1), $\Omega _{2}/\gamma =27$ (curve 2)

Fig. 4. The contour-plot of the Wigner function in the chaotic regime.

Fig. 5. Probability $P_{n}$ of finding the system in the state $\left|
n\right\rangle $ at different time intervals (curves 1 and 2) and for
regular (a) and chaotic (b) regimes. The parameters for (a) and (b)
coincides with ones for Fig. 2 curve1 and curve 2 respectively.

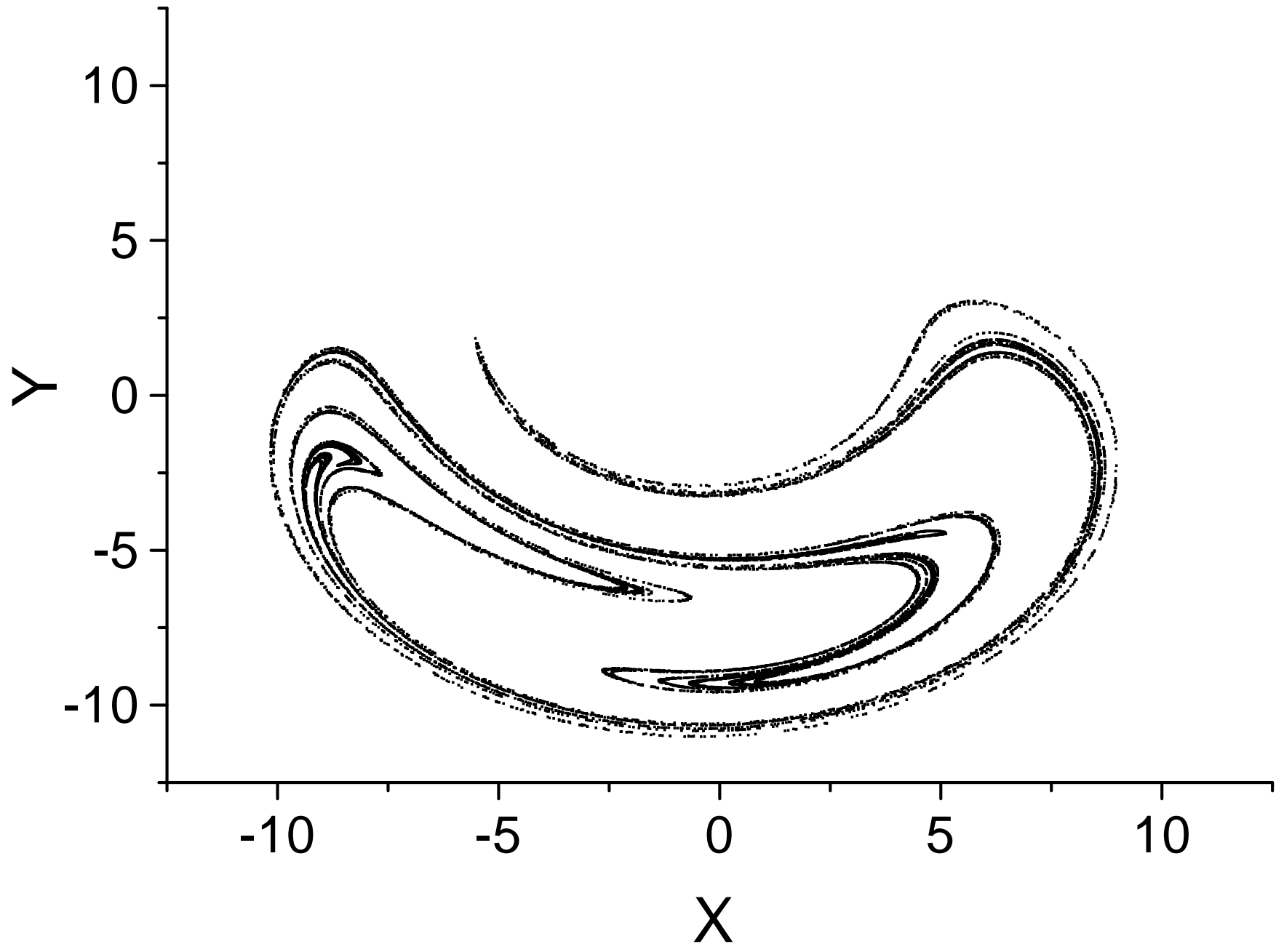
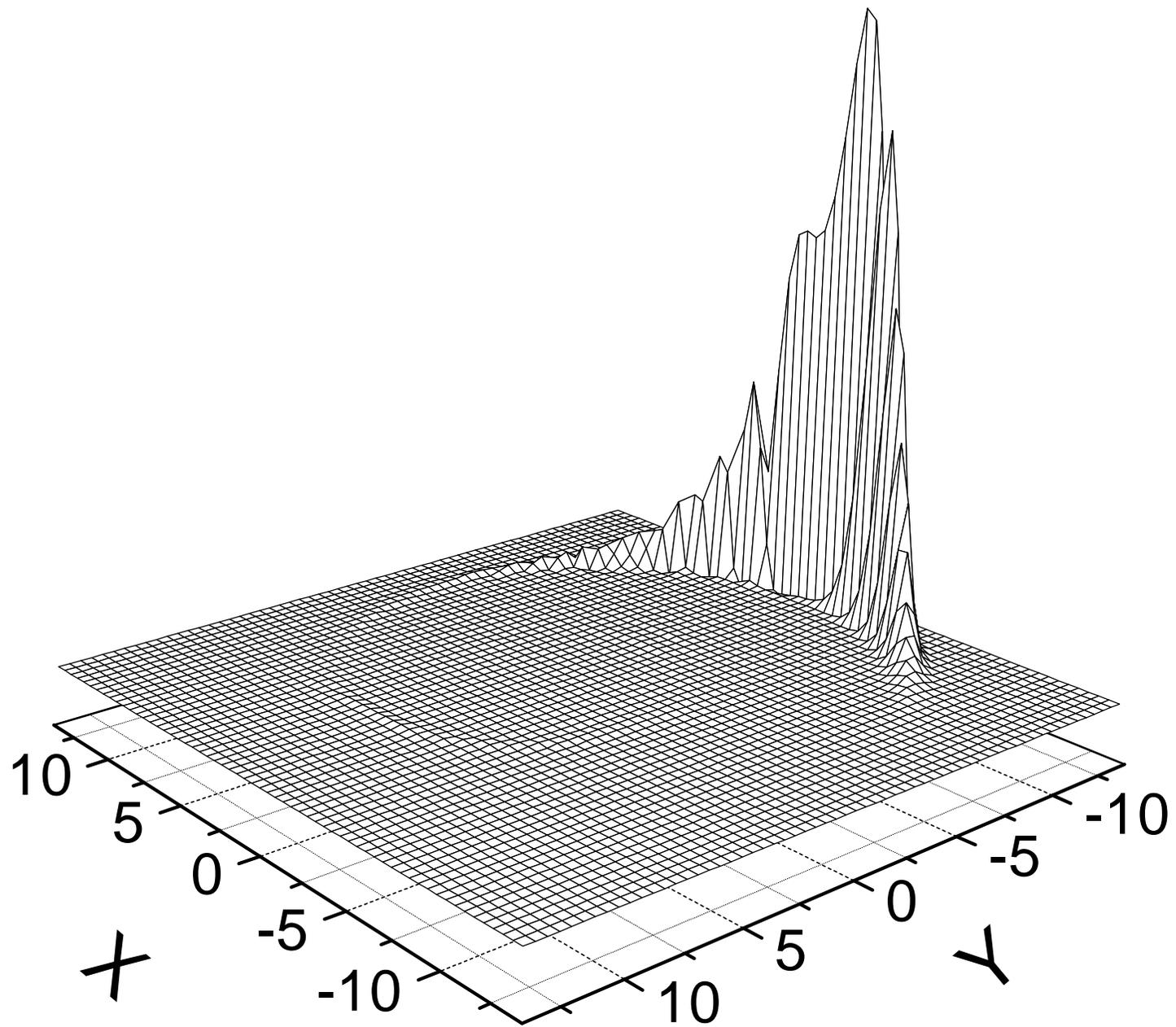
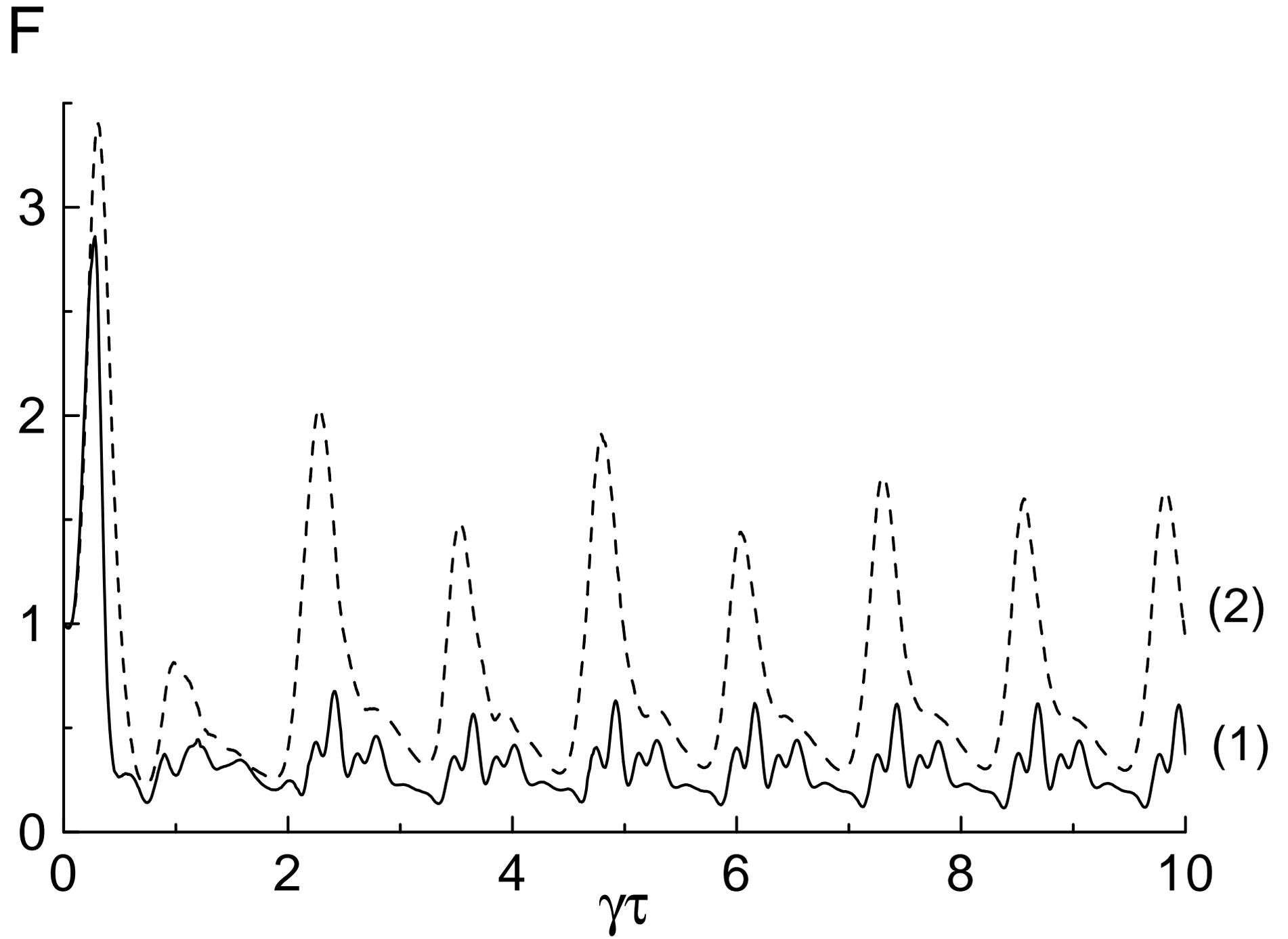
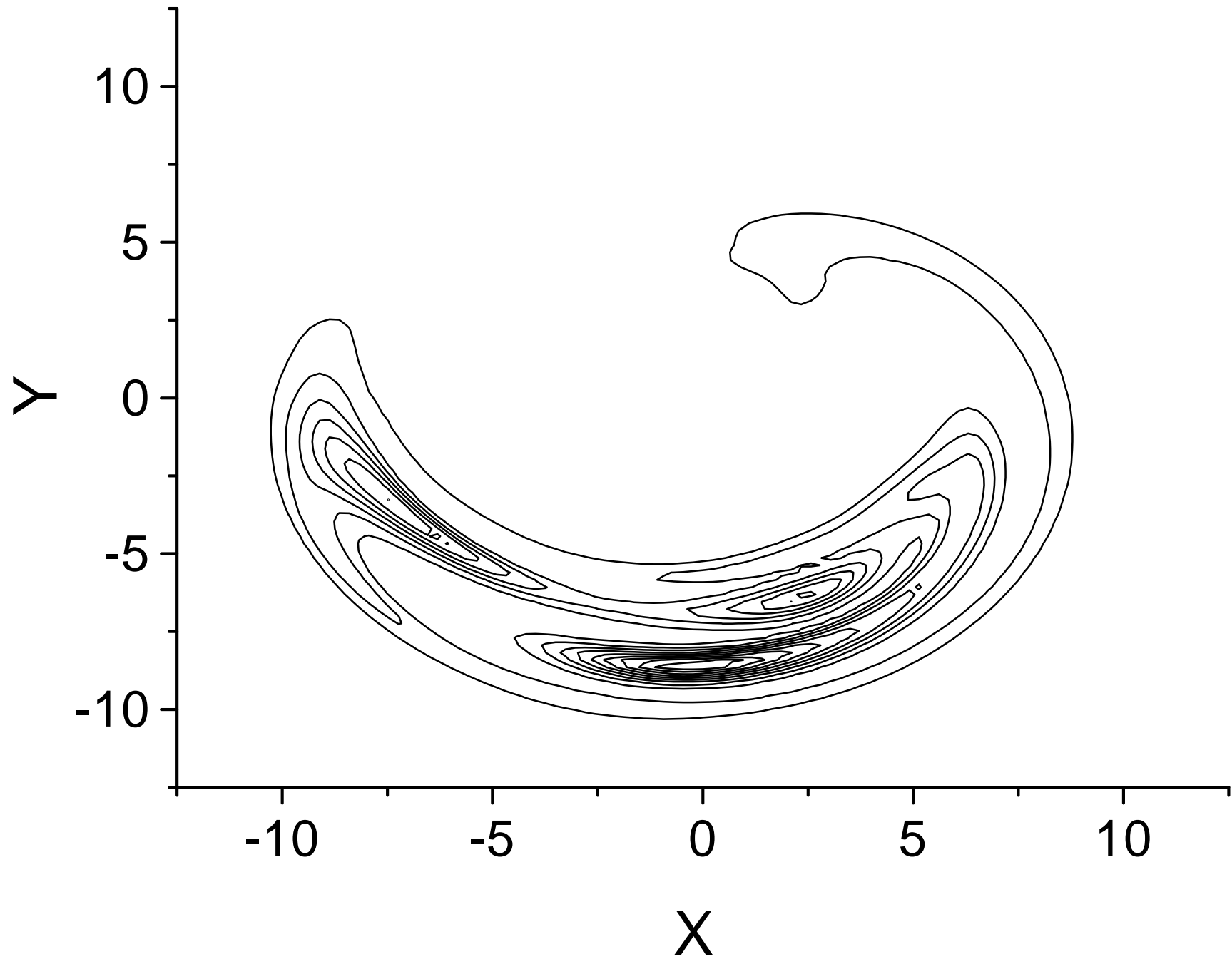
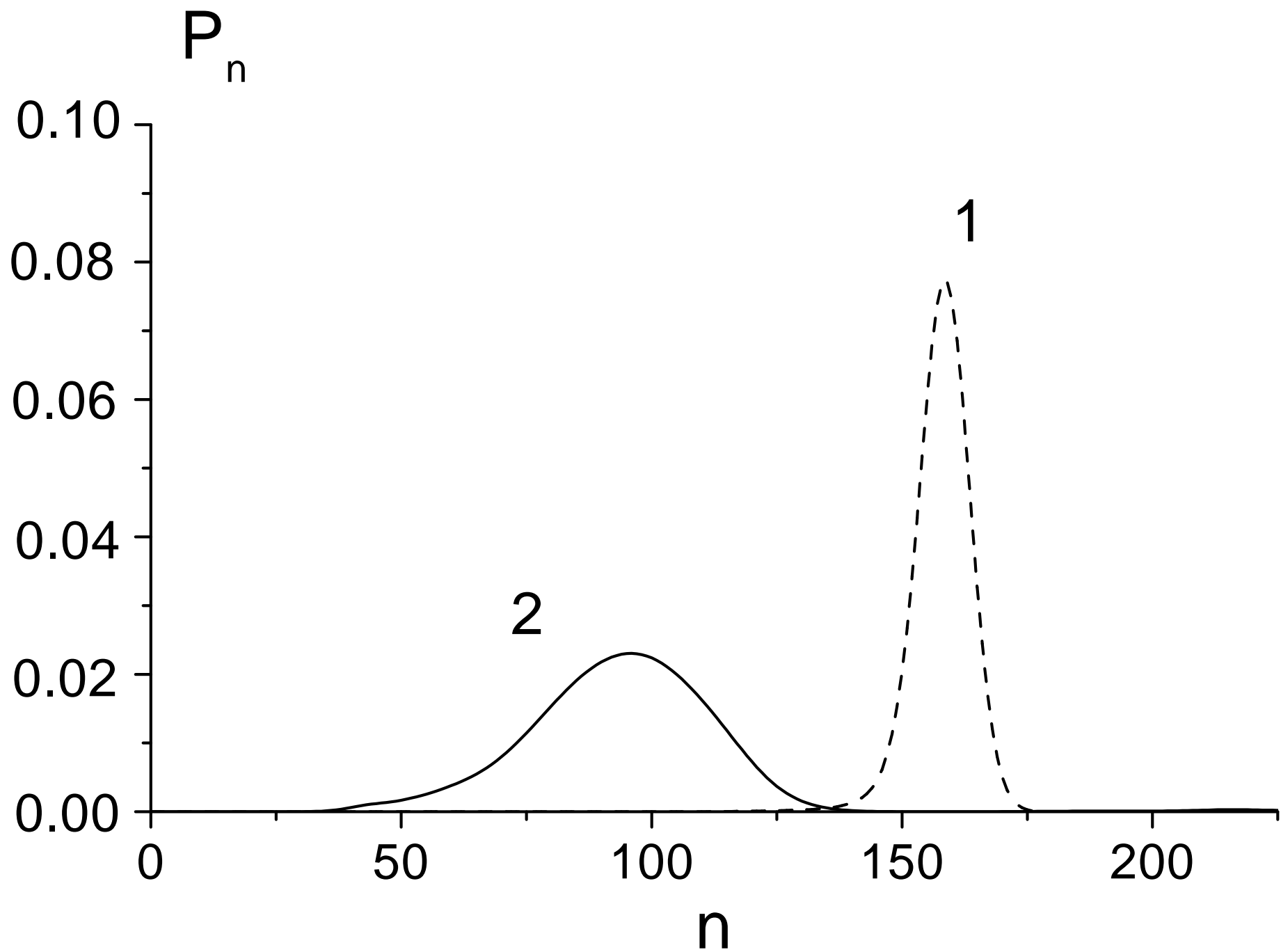
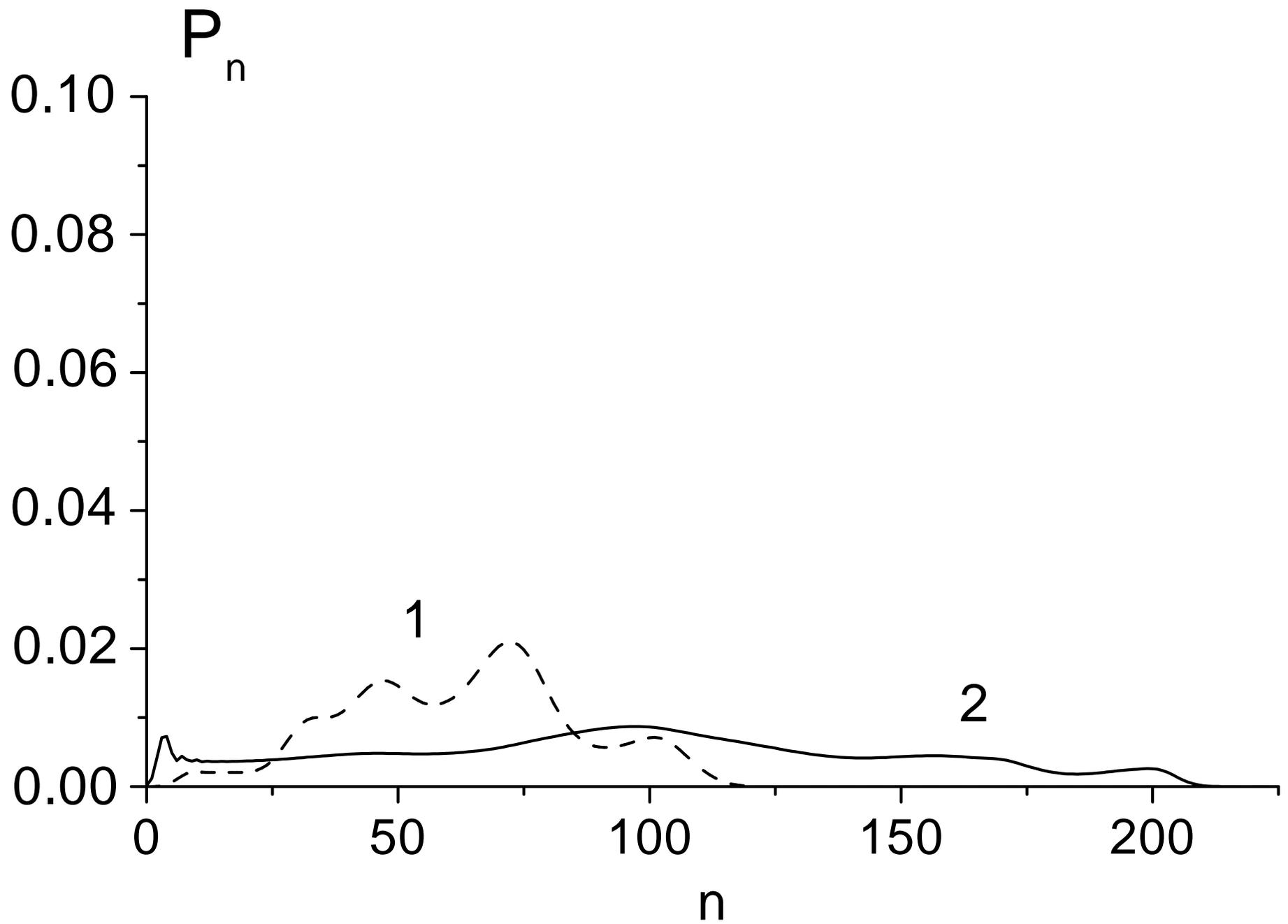
\end{document}